# Preliminary Study of a Google Home Mini

Minjin Park[1], Joshua I. James[2]

Legal Informatics and Forensic Science Institute, Hallym University, 1 Hallym University Road, Chuncheon, Gangwon 24252

[1] minjinpark819@gmail.com [2] Joshua.I.James@hallym.ac.kr

**Abstract.** Many artificial intelligence (AI) speakers have recently come to market. Beginning with Amazon Echo, many companies producing their own speaker technologies. Due to the limitations of technology, most speakers have similar functions, but the way of handling the data of each speaker is different. In the case of Amazon echo, the API of the cloud is open for any developers to develop their API. The Amazon Echo has been around for a while, and much research has been done on it. However, not much research has been done on Google Home Mini analysis for digital investigations. In this paper, we will conduct some initial research on the data storing and security methods of Google Home Mini.

## 1 Introduction

Artificial Intelligence (AI) is a critical component of the Internet of Things (IoT). The cloud connects and controls various electronic devices based on the Internet. AI can control objects according to the surrounding environment. Also, self-learning is possible through natural language understanding and deep learning, and it is tailored to the convenience of users. Meanwhile, AI is an unshaped interface, so companies combine AI with device for commercializing. That is AI speaker. The AI speaker is a speaker with an AI secretary on the Bluetooth speaker. It can work with other AI devices and can be controlled by the voice. For international AI speakers include Amazon Echo, Google's Google Home and Google Home Mini and Apple HomePod. For domestic AI speakers include SK Telecom NUGU, KT GiGA Genie, Naver Clover and Kakao Kakao Mini. The AI speaker to be covered in this paper is the Google Home Mini made by Google. The Google Home Mini is a small version of Google Home launched in October 2017 and launched in September 2018 in Korea. OS supports Android 5.0 or above, iOS 9.1 or above. To use the Google Home Mini, the user needs to sync it with the Google Home app, and the Google assistant will work when user registers and Wi-Fi certified in user app [8]. There has not been much research on the Google Home Mini compared to the Amazon Echo. In this paper, try to analyze Google Home Mini, its data storing system and security system. Research subjects include Google Home Mini, Google Home App, network, and local API. Then, with the data obtained, will be solved in terms of digital forensic perspective.

## 2 Background Research

Chung [2] presented a cloud acquisition tool for IoT environments, which is called CIFT(Cloud-based IoT Forensic). They tested it for about two months. For the quality of the test, they focused on certain products, web browsers and used python. Through the test, they got unofficial Alexa APIs, native artefacts – user accounts, Alexa-enabled devices and saved Wi-Fi settings, client-centric artefacts and web cache. Especially for the web cache for Android and Chrome, it seems as though the data has the potential to be useful digital evidence as it helped to expect the user's action. Though the paper, they focused on the findings and proof-of-concept tool to be useful for researchers who work on Amazon Alexa in digital forensic investigation view. In the paper, it will be useful to mention what kind of comments they gave, how much data left.

Clinton [12] tested Amazon echo's system on privacy and vulnerabilities for security. By tearing down the speaker, we tried to exploit the echo hardware system. The paper found three primary methods to access to the speaker, which are the SD card pinout, an eMMC style root, and JTAG. It means Amazon echo is vulnerable for the physical attack, which means people can access to the speaker and get the data out from the speaker, and the privacy of the user is also in danger. Thus, the paper suggests being aware of these kinds of matters to protect the data. Currently released AI speakers are hard to connect physically. Therefore, it would be unlikely to happen.

Wohlwend [10] discusses Amazon Alexa and echo system structure, security policy and security test for the Echo. The paper focuses on three main goals for security, which are confidentiality, integrity, and availability. Then through the four-security testing – sound, network, direct API, and third-party skills – they find that Amazon echo stores data based on the Amazon cloud server and the device resist an attack of network and API. Then imply the third-party skills can be a vulnerability. It would be good if the paper talks more detail about the API based attack and give a try on third-party skills not only mentioning on the paper.

A 2018 article [7] describes how Google voice technology works in digital forensics. 'OK, Google' can be used to unlock mobile locks. If you have only the voice file of the suspect, you can unlock the phone and extract the data. In other words, the suspect's voice can be used to find the suspect's mobile device. Also, it can be applied to other smart devices using the Google assistant. The Google Home Mini can also wake up the device with 'OK Google' or 'Hey Google', and some features require voice verification. This means that if the suspect is a user of the Google Home Mini, the suspect's voice can tell whether the device is being used or not, and if the Google Home Mini is connected to another smart device, also can find some data of the suspect.

Hyde [13] analyzed Amazon echo, and Amazon echo dot devices itself, Kasa and Alexa mobile app, network, connected devices Amazon echo has SanDisk SDIN7DP2-4G, ISP pin is out. Amazon Echo Dot has different eMMC on each board. ISP pin is also out. Through the imaging, they found Wi-Fi connections, device information log, registration information from the device. In the app, they found, some databases and in the web application, they got URL of calling and messaging. Also, for the APIs, the cards, device, Wi-Fi, Smart Home devices, Activities were found. In the Kasa, which was the TP-Link smart devices mobile app, some critical data were found such as account, password and location. They focused on what kind of data was stored and where the data was stored.

Dipert [3] explains the inside of Google Home Mini by tear down. Inside of the device, audio amplifier, metal shield with a combination of Marvell's 88DE3006-BTK2 system SoC, Toshiba NAND flash memory, two embedded antennas, SK Hynix 4GB DDR3L SDRAM, microphones and manufacturing code sticker are found. In the papers, expect the same result as this blog shows and will be a more detailed explanation.

Moore [14] investigated about Google speakers based on the several investigative questions. He analyzed Google Home app with mobile phone, device itself using software tools and chip off and lastly, cloud service. From the phone, google account information, device location, cloud device ID and wifi password are been found. By examining the internals of the speaker, open ports, GET/POST requests, and Bluetooth information that is connected to the device such as MAC address, device name and date that has been connected. From the chip off, Google Home Mini has Toshiba TC58NVG1S3HBA16 256 MB NAND flash and use BGA 67 Socket, NAND Flash Chip Reader (Dataman) to read the data inside of the chip. From the dump, Bluetooth device information and Google account. For cloud acquisition, mainly used Google's 'My Activity' and Google Takeout. Through acquisition, not all but some given commands and answers were found. for the future work, parse Google Home file system, decoding of the proto file, iOS app examination and calling feature.

## 3 Research Problem and Methodology

To date, very few works have looked at the Google Home Mini (GHM) from a digital investigation perspective. This work is an initial analysis of the locations and types of data related to the Google Home Mini, and is meant to be a starting point for future data acquisition studies. In this work, we separate the study into three main sections: the device, the mobile app and the network. Based on our past experiences, each of these locations provide different – but related – data that may be of evidential value.

We used Google Home Mini's built-in functions and direct interaction to generate user-related data. Commands were given directly to the Google Home Mini (serial number: 7B28L5NRWF) as well as via the mobile phone application (Galaxy note 4, SM-N910S, Android version 6.0.1). For this research, Korean and English were used to give commands. After the function test, chip-off was used to acquire additional embedded device data. After the chip off, the chip was imaged using Hancom GMD MD-series acquisition software.

Google has a mobile app called 'Google home'. The app version we used was 2.9.40.16. A user can give a command from the phone by speaking or typing. Also, the user can control the device in the app. To extract the data that is stored in the app, the ES File Explorer has been used. ES File Explorer is a free and featured file (application, document and multimedia) manager for local and network use. The version is 4.1.9.9.21. For the analysis, the phone (Galaxy Note4, SM-N910S) and laptop (Samsung, NT905S3G-KSQB) have been used. First, it generates data using the app. Then in the ES File Explorer, check whether the phone is rooted. Second, follow the path: device/data/data/Chromecast. Third, copy the Chromecast folder and paste in the internal storage. Then connect the phone to the laptop using a USB cable to extract the Chromcast folder. After all these steps, in this research, use Autopsy (version 4.10) to analyze the data. Add folder as Logical file.

In the network analysis, use Wireshark (version 3.0.0) and Zenmap (version7.7.0) to acquire the data. Figure 1. shows how the network has connected for the test. Google Home Mini and the mobile phone are connected to the laptop - the laptop works as an access point using its hotspot - and the laptop is connected to the internet and internet connects to the Google cloud.

We use the GUI version of Nmap, Zenmap, to look for open ports in the speaker's IP address. Use Telnet to see if the ports are listening. The test the vulnerability according to the usage of each port. Vulnerability testing uses Metasploit, Heartbleed and Openssl with Kali Linux. After that, uses Burp suite (v1.7.36) and Postman to acquire the data. In the Burp Suite, under proxy, add the laptop IP address and port 8080 and in the mobile phone Wi-Fi advanced setting, set proxy manual and port 8080. Then it intercepts the network of the speaker that talks to the phone through the laptop.

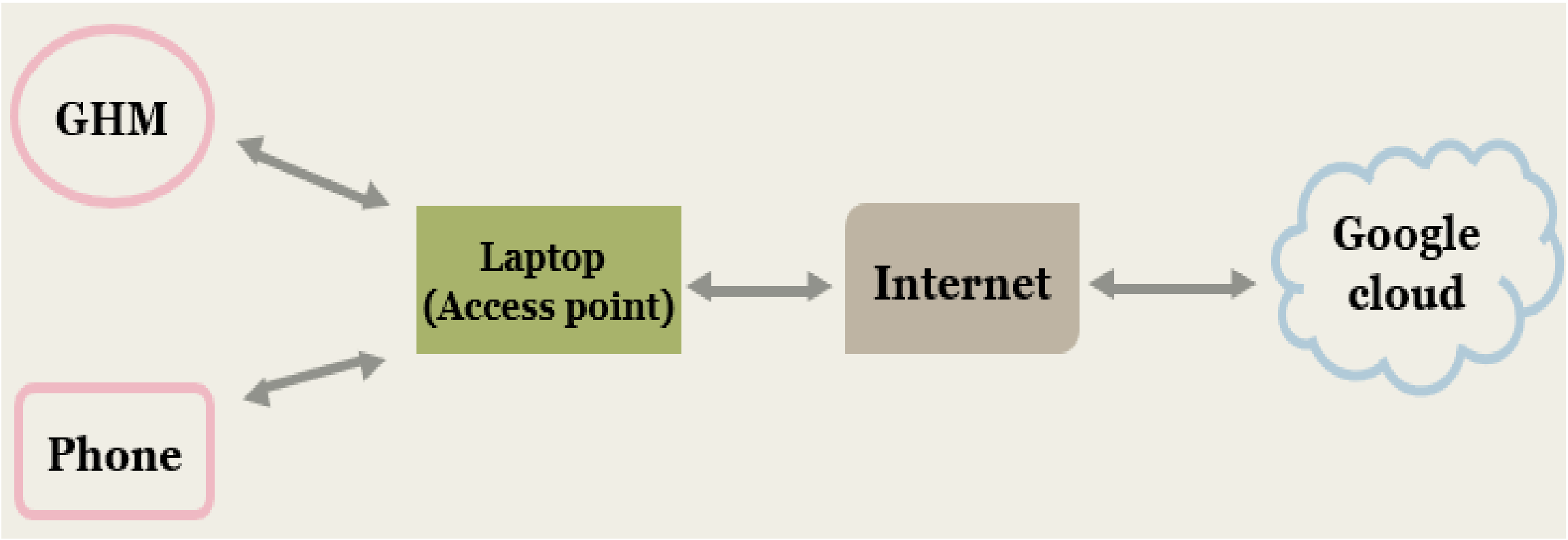

**Fig. 1.** Google Home Mini network setup for device, mobile app and network data collection.

# 4 Data Collection and Analysis

This section covers different types of analysis we conducted on the Google Home Mini hardware and software with a focus on the mobile app analysis.

## 4.1 Google Home Mini Hardware

The Google Home Mini measures 98mm in diameter, 42mm in height and weight 173g. The colours are chalk, charcoal, coral-google store exclusive, and durable fabric material on the top. The Wi-Fi network supports 802.11b/g/n/ac (2.4 GHz/5 GHz) and features 4.1 Bluetooth. There is a microphone on/off switch and a micro-USB power connector on the back and a reset button right below the power connector.

## 4.2 Google Home Mini Functions

To use the functions, the user should say wake-up words are ‘Hey, google’ and ‘Ok, google’. About 20 languages are available. If the user talks in English, the speaker responses in English. If the user talks in Korean, the speaker responses in Korean. Also, it provides information related to the language. There are many functions. It is similar to other speakers — for example, weather, alarm, news, traffic, and more. Among the functions, there are two functions which are ‘Routine’ and ‘Remembering Things’. User can make a routine command to suit their lifestyle. For example, if the user says ‘Bedtime, the speaker tells tomorrow’s weather, tomorrow’s first calendar event and ask what time the alarm should be set for then play sleep sounds. It will be a good way to know user behaviour. For ‘Remembering Things’, once the user gives the command ‘remember that my key is on the desk next to the door’, Google Home Mini accept the command, and when a user asks for the key, it tells where the key is. It will be useful to find something about the user.

## 4.3 App Analysis

Table 1 shows data and locations from the Google Home app on the mobile phone. Some tokens, account ID, the nickname of the device, address of the speaker, app version, port, phone IP address and Wi-Fi name and password list. For the Wi-Fi and password list, the phone has been connected to three different networks. Neo_house6 is the actual network router, and DESKTOP-ENIL7DS is a hotspot from the laptop. Both laptops were connected to the router and hotspot.

With the information from the app, it can be used when the users deny that they have never use the speaker or they do not have one. The user ID and location will verify and identify where the user is and was. Also, with the tokens, extract the data from the cloud is available. Also, Wi-Fi tells the connection between the user and the speaker.

**Table 1.** Potential evidential data found from the Google Home Mini Mobile App Analysis

| Location | Data |
|---|---|
| /LogicalFileSet1/com.google.android.apps.chromecast.app/ shared_prefs/com.google.android.gms.appid.xml | • LastToken<br>• appVersion |
| /LogicalFileSet1/com.google.android.apps.chromecast.app/ shared_prefs/ com.google.android.apps.chromecast.app_preferences_no_backup.xml | • lastRefreshTime<br>• selected_routine_device_id<br>• ph_server_token<br>• gcmIdToken |
| | • 61 Sakju-ro, Gyo-dong, Chuncheon, Gangwon-do, South Korea<br>• Longitude/ latitude<br>• home_graph_last_refreshed_simonhallym@gmail.com<br>• addressLine2 Chuncheon, Gangwon-do 200-060<br>• current_account_name simonhallym@gmail.com<br>• Wi-Fi name and password list [{"n":"neo_house6","p":"********","s":2},{"n":"DESKTOP-ENIL7DS 3926","p":"dkssudgktpdy!!","s":1}, {"n":"me","p":"********","s":2}]<br>• current_home_id_simonhallym@gmail.com<br>• addressLine 61 Gyo-dong |
| /LogicalFileSet1/com.google.android.apps.chromecast.app/shared_prefs/com.google.android.apps.chromecast.app_preferences.xml | • live_card_consistency_token com.google.h.b.d.a.w@7bc6f<br>• dismissedActionChipSetupDevicesFA:8F:CA:98:A5:5<br>• setup-salt e3452b4b-9fd6-42b6-8e47-e71bc8dd0741 |
| /LogicalFileSet1/com.google.android.apps.chromecast.app/cache/cronet_http_cache/prefs/local_prefs.json | • servers:https://googlehomefoyer-pa.googleapis.com<br>• expiration:13200914931582426<br>• port:443<br>• address:192.168.166.11 |
| /LogicalFileSet1/com.google.android.apps.chromecast.app/files/home_graph_c2ltb25oYWxseW1AZ21haWwuY29t.proto | • OfficeZeLIFS<br>simonhallym@gmail.com<br>216-33 Gyo-dong, Chuncheon, Gangwon-do, South Korea<br>• bettyhallym@gmail.com2<br>• google.com:api-project-498579633514 5759C8B0CEAFB4B8D438569D3288716F: 41d28897-d2b4-4d80-bc26-de7057ec36b6<br>• google.com:api-project-498579633514 5759C8B0CEAFB4B8D438569D3288716F |

### 4.4 Network Analysis

We used Wireshark to capture the network traffic between the access point, the Google Home Mini and the mobile device running the GHM app.

| | | | | | |
|---|---|---|---|---|---|
| 739 | 249.531437 | googleapis.l.google.com | 192.168.137.37 | TLSv1.2 | 362 Application Data |
| 740 | 249.531654 | googleapis.l.google.com | 192.168.137.37 | TLSv1.2 | 188 Application Data |
| 741 | 249.531786 | googleapis.l.google.com | 192.168.137.37 | TLSv1.2 | 238 Application Data |
| 742 | 249.541653 | 192.168.137.37 | googleapis.l.google… | TCP | 66 55024 → 443 [ACK] Seq=1715 Ack=3975 Win=96256 Len=0 TSval=349262 TSecr=3017774295 |
| 743 | 249.567995 | 192.168.137.37 | googleapis.l.google… | TLSv1.2 | 104 Application Data |
| 744 | 249.578871 | 192.168.137.37 | googleapis.l.google… | TCP | 1434 55024 → 443 [ACK] Seq=1753 Ack=3975 Win=96256 Len=1368 TSval=349269 TSecr=3017774295 |
| 745 | 249.579025 | 192.168.137.37 | googleapis.l.google… | TLSv1.2 | 397 Application Data |

**Fig. 2.** Network traffic collection between the mobile device (app) and the Google Home Mini cloud service.

Figure 2 shows the network packet between the mobile and the cloud. During this time, the commands have been given to the Google Home mobile application. As Figure 3 shows, Google APIs and mobile share the application data with TLSv1.2, which is encrypted. Also, sometimes use the latest encryption protocol TLSv1.3 encryption as well [1].

| | | | | |
|---|---|---|---|---|
| 11.566700 | 192.168.137.37 | googleapis.l.google.com | TLSv1.3 | 583 Client Hello |
| 11.728772 | googleapis.l.google.com | 192.168.137.37 | TLSv1.3 | 2954 Server Hello, Change Cipher Spec |
| 11.734361 | googleapis.l.google.com | 192.168.137.37 | TLSv1.3 | 2288 Application Data |
| 11.746815 | 192.168.137.37 | ytimg-edge-static.l.go… | TLSv1.3 | 583 Client Hello |
| 11.839652 | 192.168.137.37 | googleapis.l.google.com | TLSv1.3 | 130 Change Cipher Spec, Application Data |
| 11.912633 | ytimg-edge-static.l.google.com | 192.168.137.37 | TLSv1.3 | 2954 Server Hello, Change Cipher Spec |
| 11.920851 | ytimg-edge-static.l.google.com | 192.168.137.37 | TLSv1.3 | 572 Application Data |
| 11.924652 | 192.168.137.37 | ytimg-edge-static.l.go… | TLSv1.3 | 130 Change Cipher Spec, Application Data |
| 11.985559 | googleapis.l.google.com | 192.168.137.37 | TLSv1.3 | 1122 Application Data |

**Fig. 3.** Network traffic collection between the mobile device (app) and the Google Home Mini cloud service using TLSv1.3

| | | | | | |
|---|---|---|---|---|---|
| 390 | 32.291502 | c812a7f8-e657-5590-d4c0-6f1b40… | www.google.com | TLSv1.2 | 212 Application Data |
| 391 | 32.291589 | c812a7f8-e657-5590-d4c0-6f1b40… | www.google.com | TCP | 1434 46853 → 443 [ACK] |
| 392 | 32.291590 | c812a7f8-e657-5590-d4c0-6f1b40… | www.google.com | TLSv1.2 | 155 Application Data |
| 393 | 32.307089 | www.google.com | c812a7f8-e657-5590-d4c… | TLSv1.2 | 632 Application Data |
| 394 | 32.307305 | www.google.com | c812a7f8-e657-5590-d4c… | TLSv1.2 | 666 Application Data |

**Fig. 4.** Network traffic collection between the Google Home Mini speaker (device) and the Google cloud service.

Figure 4 is the network packet between the Google Home Mini and google browser. It is captured when the command has given to the device. As Figure 4 shows, it also uses the TLSv1.2 encryption to share the application data.

After finding encrypted network traffic using Wireshark, we used Zenmap to discover the open ports as a first step to detect any vulnerabilities. The test takes the same setup condition as Wireshark. There were five open ports for the Google Home Mini in TCP protocol, and it is also the same for UDP protocol. In the previous research [14] also showed same number of the ports so it seemed to be fixed open ports. Among these ports, the test focuses on two ports, which are 8009 and 8443. We used Telnet to test open ports for listening status and service detection. Through Telnet testing, all five ports were found to be listening. Then, we investigated those ports with Kali Linux. For the test, without using hotspot of the laptop, put speaker, mobile phone and laptop in the same network, which is neo_house6.

Basic known attacks were conducted against the open ports using Metasploit. No ports were immediately vulnerable to known attacks. Next, we moved to a local API test.

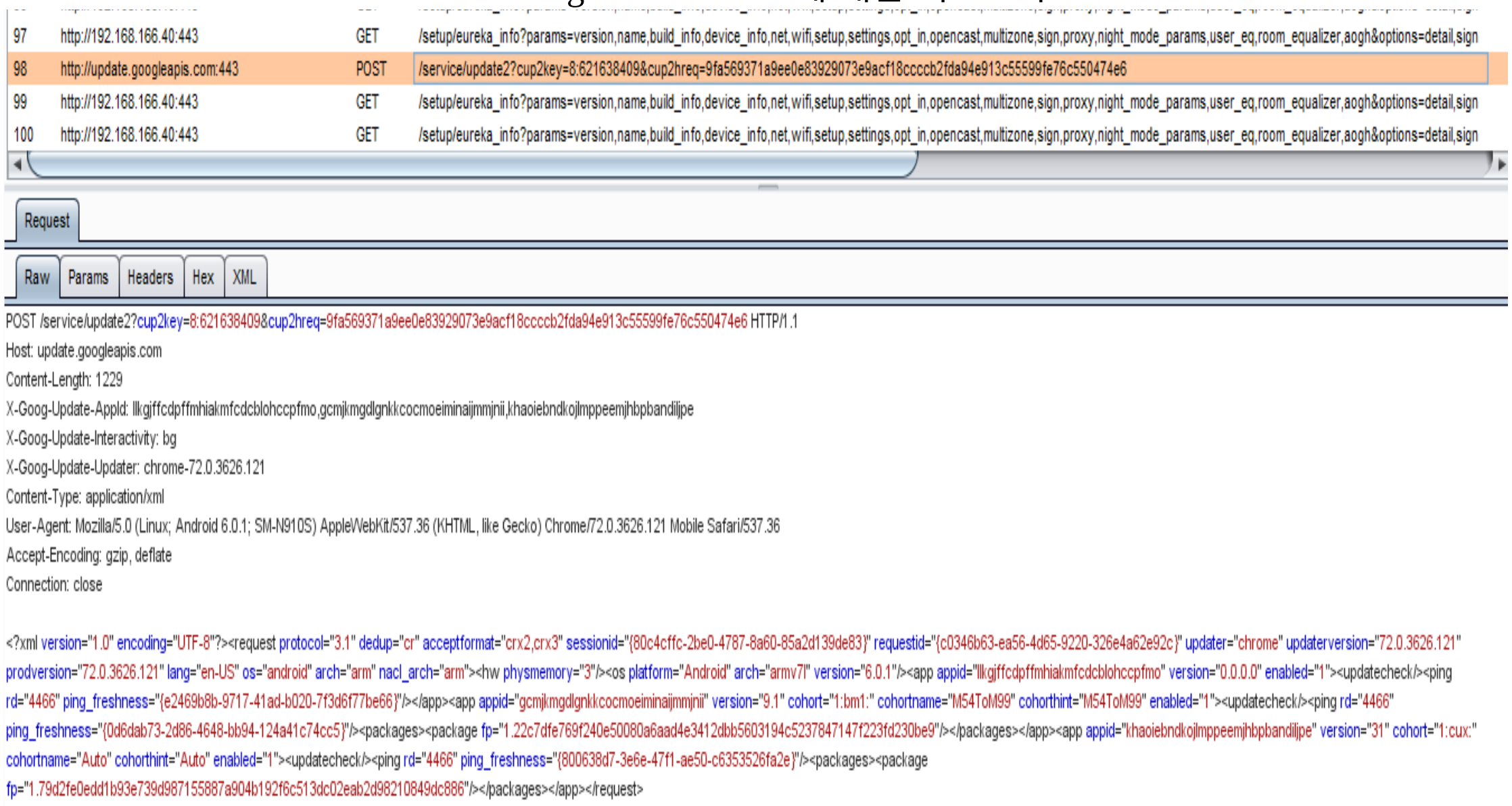

**Fig. 5.** Some of APIs are found by using Burp Suite.

By using Burp Suite, some APIs are found which are the app_device_id and parameters for version, name, build_info, device_info, net, wifi, setup, settings, opt_in, opencast, multizone, sign, proxy, night_mode_params, user_eq, room_equalizer, aogh&options in figure 5. Also, aglio [5] has created an unofficial APIs that is used between the Google Home app and Google Home so apply some APIs to the Google Home Mini to find the APIs. Findings are in Table 2.

**Table 2.** API calls and information returned from a Google Home Mini on the local network.

| Type | Info | API | Returned |
|---|---|---|---|
| GET | Parameters | http://192.168.166.40:8008/ setup/eureka_info? params=version,name,build_info, device_info,net,wifi,setup,settings,opt_in,opencast,multizone,sign, proxy,night_mode_params,user_eq,room_equalizer,aogh&options =detail,sign | device_info:<br>cloud_device_id: "D2C293358C936F117 57914443A7C3F57",<br>factory_country_code: "US",<br>hotspot_bssid: "FA:8F:CA:98:A5:5B",<br>mac_address: "20:DF:B9:4E:87:FE",<br>manufacturer: "Google Inc.",<br>model_name: "Google Home Mini",<br>product_name: "mushroom",<br>name: "Office speaker",<br>net: ip_address: "192.168.166.40",<br>settings:<br>country_code: "KR",<br>locale: "en-US",<br>timezone: "Asia/Seoul",<br>wifi:<br>bssid: "90:9f:33:db:10:de",<br>ssid: "neo_house6" |
| GET | Eureka Info | http://192.168.166.40:8008/<br>setup/eureka_info | bssid: "90:9f:33:db:10:de",<br>hotspot_bssid: "FA:8F:CA:98:A5:5B",<br>ip_address: "192.168.166.40",<br>locale: "en-US",<br>location:<br>country_code: "KR",<br>latitude: 255,<br>longitude: 255<br>mac_address: "20:DF:B9:4E:87:FE",<br>name: "Living Room",<br>ssid: "neo_house6",<br>timezone: "Asia/Seoul",<br>uma_client_id: "cc918aa3-2bba-46d2-aa3c-bfe1fa3c275b" |
| GET | offer | http://192.168.166.40:8008/<br>setup/offer | token: "ADtqmfTJx82eFvi_wg3BOUfBcUmZgF_ik7veTnYR0hc9MTyJxXQZIJb_OY4B2CEvZizrabJqcZp4DyjvCPBV53Ya1qJ05SdNiY5zxADnaIB04sSiflT-IjZUj2yaowZFQxlQUFHLiKDm" |
| GET | Time zone | http://192.168.166.40:8008/<br>setup/supported_timezones | display_string: "Hawaii-Aleutian Standard Time (Honolulu)",<br>timezone: "Pacific/Honolulu"<br>display_string: "Hawaii-Aleutian Daylight<br>*support many different time zone |
| GET | Supported locales | http://192.168.166.40:8008/<br>setup/supported_locales | display_string: "Amharic - አማርኛ",<br>locale: "am"<br>display_string: "Arabic - العربية",<br>locale: "ar"<br>*support many different locales |
| GET | Alarm | http://192.168.166.40:8008/<br>setup/assistant/alarms | day: 1,<br>month: 4,<br>year: 2019<br>fire_time: 1554108270000,<br>id: "alarm/5d762a93-0000-20b9-9fa8-f4f5e80b89c8",<br>status: 1,<br>time_pattern:<br>hour: 17,<br>minute: 44,<br>second: 30 |
| GET | Bluetooth | http://192.168.166.40:8008/<br>setup/bluetooth/status | connected_devices:<br>device_class: 5898764,<br>mac_address: "10:92:66:13:c0:4a",<br>name: "Hallym Simon (Galaxy Note4)" |
| GET | Configured network | http://192.168.166.40:8008/<br>setup/configured_networks | ssid: "me",<br>ssid: "DESKTOP-ENIL7DS 3926",<br>ssid: "neo_house6" |
| POST | App device ID | http://192.168.166.40:8008/<br>setup/get_app_device_id | "app_device_id":<br>"D2C293358C936F11757914443A7C3F57" |
| POST | Internet download speed | http://192.168.166.40:8008/<br>setup/<br>test_internet_download_speed | "bytes_received": 31457280,<br>"response_code": 200,<br>"time_for_data_fetch": 21807,<br>"time_for_http_response": 819 |
| POST | Scan Wi-Fi | http://192.168.166.40:8008/ setup/scan_wifi | Response 200 |
| GET | Scan results | http://192.168.166.40:8008/ setup/scan_results | bssid: "90:9f:33:db:10:de",<br>bssid: "90:9f:33:db:10:de",<br>ssid: "neo_house6" |

The data can be used for investigation. For example, tokens can be a key to extract the data from the cloud. In parameters and some APIs have the network information of the speaker, which can tell which network the speaker is using. Other AI products connected through this network can also be identified. If the Bluetooth function is on, the investigator can check the information of the device connected to the speaker. If the speaker is connected to the mobile, it can discover the user's mobile model and MAC address.

### 4.5 Chip-off Analysis

After imaging the chip off data, a small file of 90,213KB was created and it was all in carved files. The carved files are analyzed with Autopsy 4.10.0.

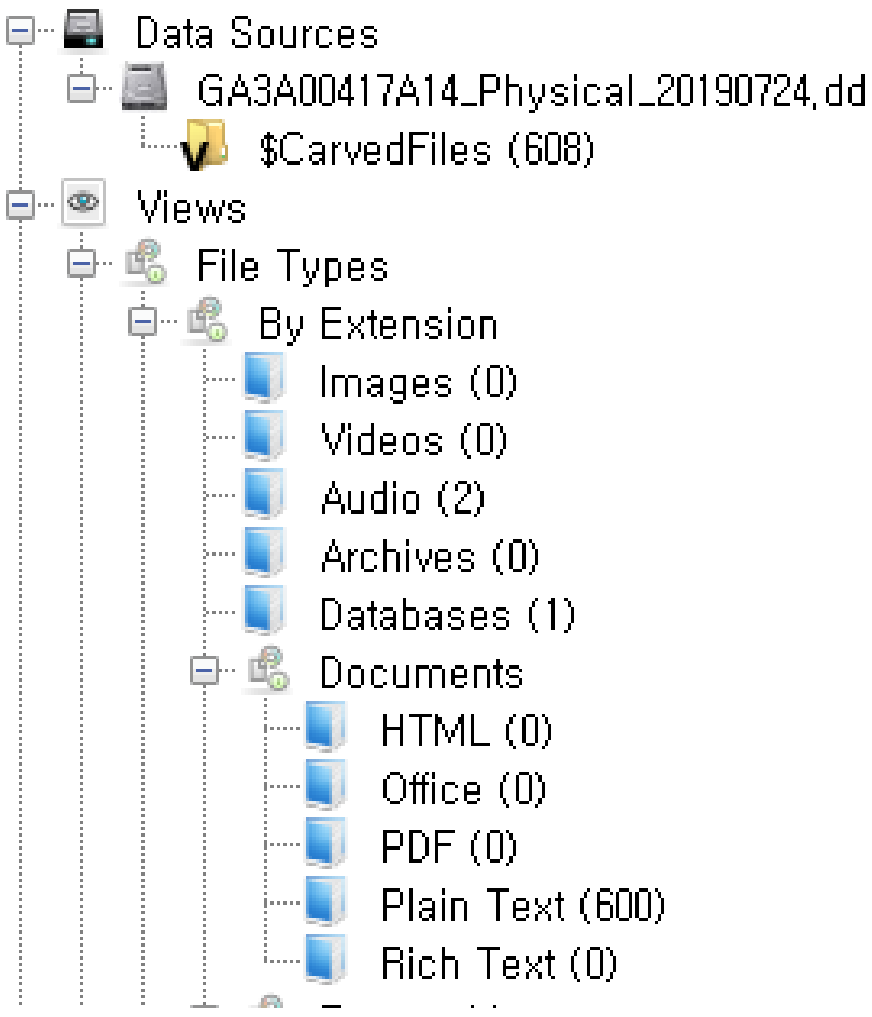

**Fig. 6.** .dd file opened with Autopsy to analyze

Bunch of log text files are stored but it does not contain much information. Table 3 shows some findings.

**Table 3.** Acquired log text data from chip-off

| File name | Data |
|---|---|
| F0161412 | • RAM: 476992K total, 271004K free, 42636K buffers, 83604K cached, 108K shmem, 9252K slab<br>• Kernel log |
| F0166284 | • NAND device: Manufacturer ID: 0x98, Chip ID: 0xda (Toshiba 256MiB 8-bit) |
| F0168388 | • Product name: mushroom<br>Product model: Google Home Mini |
| | • Wifi.interface: mlan0<br>Wifi.ap.dev_name: ipTIMEAP<br>Wifi.ap.manufacturer: ipTIME<br>Wifi.ap.model.name: ipTIMEAP<br>Wifi.ap.modle_number: 1234567890<br>Wifi.ap.vendor_prefix: [90:9f:33] |
| F0167416 | • OS platform: Linux |
| F160746 | • User name: Hallym simon<br>Phone model: Galaxy Note 4<br>MAC address: [10:92:66:13:c0:4a] |

The .ogg files were open with Groove 2018 from Microsoft and VLC media player (version 3.0.7.1) but all failed. Even converted ogg to mp3 file, did not work. In addition, the sqlite file was also not able to open. It is expected that corruption would have occurred during the imaging process.

## 5 Conclusion

Through the test, several facts are confirmed. First, Google Home Mini does not store much data in the app. Second, by using Wireshark, it has found that how Google Home Mini, Google Home app and the Google cloud exchange data. They exchange the data using TLSv1.2 encryption and occasionally using the latest version of encryption TLSv1.3. Third, Google Home Mini has five open ports, which are HTTP, ajp13, https-alt, cslistener and scp-config. The vulnerability tests on the ajp13 and https-alt ports are done, and they are not exploitable. Then use the Burp suite and Postman to find the local APIs. In the API, some essential data exists, like tokens that can be used for cloud data, configured network information of speakers, Wi-Fi information, setting country for speaker, time zone, and Bluetooth information. For chip off, not much data is stored These data can be applied if the suspect has a Google Home Mini in his house, and the suspect is denied telling its use and connection between them. Additional work still needs to be done on the device, application and network levels of the Google home mini to find additional, potentially hidden, data. Further, software updates for the GHM are frequent, and new artefacts may be introduced with each new update. Also, this test has don only for Android. Thus, the test for iOS will be needed to get more date about Google Home devices.

**Acknowledgments.** This work is funded by the KIAT (Korea Institute for Advancement of Technology) grant number N0002260. Thank you to Hancom GMD for support on the project.